\documentclass[prb,aps,preprint,superscriptaddress]{revtex4}
\usepackage{graphicx}
\usepackage{amssymb}
\usepackage{bm}

\begin{document}

\author{Toshifumi Itakura}
 \affiliation{ASPRO TEC., Takabaridai 3-506, Meito-ku, Nagoya 465-0054, Japan}
 \email{itakurat@s6.dion.ne.jp}

\title{Decoherence of coupled spin qubit system}

\begin{abstract}
In this study, we examine decoherence of a spin qubits system coupled independently to counted spin chain with a $1/r^2$ interaction by using influence functional. We also examine the time evolution of density matrix numerically when environment is Gaussian noise.
\end{abstract}


\maketitle

\section{Introduction}
Among various proposals for quantum computations, quantum bits (qubits) in solid materials, such as superconducting Josephson junctions
\cite{Nakamura},
           and quantum dots
\cite{Hayashi,Tanamoto,Loss},
           have the advantage of  scalability.
Such coherent two level systems constitute qubits and a quantum computation can be carried out as a unitary operation applied to many qubit systems.
It is essential that this quantum coherence be maintained during computation.
However, dephasing is hard to avoid, due to the interaction between qubit system the environment.
The decay of off-diagonal elements of the qubit density matrix signals occurrence of dephasing.
Various environments can cause dephasing.
In solid systems, the effect of phonons is important
\cite{Fujisawa_SC}.
The effect of electromagnetic fluctuation has been extensively studied for Josephson junction charge qubits
\cite{Schon}.

For spin qubit system, the fluctuations of the nuclear spins of impurities can also be a cause of dephasing.
It has recently been shown experimentally that the coupling between the spin of an electron in a quantum dot and the environment is very weak
                  \cite{Fujisawa_Nature,Erlingsson,Khaetskii}.
Decoherence of nuclear spins is experimentally examined, in which the source of decoherence is dipole-dipole interaction.
     \cite{Yusa}
For this reason, the dephasing time of a spin qubit conjectured to be very long.
However, both donor impurities and nuclear spins in semiconductors
       \cite{Ladd}
       have been suggested as possible building blocks for feasible quantum dots architectures.
The proposal based on experimental findings of quantum computation by using 
 an Si$^{29}$ array is another possibility
   \cite{Ladd}.
For a spin qubit system in 2D, the coupling with neighbor spin chain can cause a dephasing.
When an interaction is between the qubits themselves, it can in principle be incorporated into the quantum computer Hamiltonian, although this would lead to more complicated gate sequences.
Therefore it is instructive to analyze the error introduced by ignoring some of these interactions, as done in the case of dipolar coupled spin qubits
   \cite{Sousa_1}
   and spin-orbit interaction.
   \cite{Burkard}
For a quantum spin chain with a $1/r^2$ interaction, an exact expression of the dynamical correlation function has been obtained
\cite{Haldane}.
Using this expression, we consider the case in which the qubit system is coupled to spin of the spin chain. 
We examine the relaxation phenomena of a spin qubits array which is coupled to a spin chain with long-range interactions.
Integrating over the spin chain variables, we obtain the influence functional of the qubit system.
\cite{Weiss,Itakura}
Using this influence functional, we examine the dephasing of the qubits density matrix.
In the present study, we especially concentrated on the effect of spin flip process.
We examine the zero-dimensional qubit and one-dimensional qubits system.
The spin flips process lead to oscillation self-excitation.
\cite{Tokura_s}

\section{Hamiltonian}
We examine the Hamiltonian of counted spin chain and qubits system. 
The type of pin chain and qubit interaction is XXZ,
($A_{\perp}, A_{zz} \ne 0$).
\begin{eqnarray}
  H_{spin} &=& J \sum_{n,m} (-1)^{n-m} [d(n-m)]^2
   {\bf \vec{S}}_n \cdot \bf{ \vec{S}}_m, \\
  H_{qb} &=& \sum_i \hbar \omega_o I_i^z, \\
 H_{int} &=& \sum_i \gamma_N \hbar^2 [
\frac{1}{2} A_{\perp} (S_{i}^x I^x_i + S_{i}^y I^y_i ) 
 + A_{zz} S_{i}^z I^z_i ]. \nonumber \\
\end{eqnarray}
This Hamiltonian $H_{spin}$ represent inverse square interacting spin chain 
system, and is counted Haldane-Shastry model.
Here, $d(n)$=$(N/\pi) \sin (\pi n /N)$,
$S_{i}^{j}$,($j=x,y,z$) is spin operator at site $i$ of spin chain,
$I^{j}_i$,($j=x,y,z$) is qubit operator at site $i$,
$J$ is strength of interaction of spin chain system,
$\gamma_N$ is geomagnetic ratio of qubit.
Above model, the spin chain and zero-dimensional 
qubit interacts with contact interaction.
We also examine the one-dimensional qubit system with contact interaction,
for this case the effect of indirect interaction appears.
And, we scale the length by the lattice length $a$.

\section{single qubit system}

First, we consider a single qubit and one-dimensional spin chain system.
Thus, we examine the effect of direct interaction by influence functional method.
\cite{Weiss,Itakura}
The interaction Hamiltonian is as below,
\begin{equation}
  H_{int} = \gamma_N \hbar^2 [
   \frac{1}{2} A_{\perp} (S_{0}^x I^x + S_{0}^y I^y ) 
 + A_{zz} S_0^z I^z ].
\end{equation}

We integrate about the qubit system except the 0-th site spin.
This lead density matrix of spin system,
\begin{equation}
 \rho ( I_{z+}^f, I_{z-}^f )
   =\int^{I_{z+}(t) =I_{z+}^f, I_{z-} (t) =
    I_{z-}^f}_{ I_{z+}(0)=I_{z+}^f, I_{z+} (0) = I_{z+}^f} 
 [ d I_{z+} ] [ d I_{z-} ] 
  {\exp} ( \frac{i}{\hbar} (I_{qb} [ I_{z-} ] - I_{qb} [ I_{z+} ] ))
  F[I_{z+},I_{z-}], 
\end{equation}
the influence functional is
\begin{eqnarray}
 F[ I_{z+}, I_{z-} ] 
 &=& \int [ d  {\bf S}_{+i}  ] [ d  {\bf S}_{-i} ] \nonumber \\
 \delta (  {\bf S}_{+i} (t) - {\bf  S}_{-i}(t) ) 
 \rho (  {\bf S}_i  (0),  {\bf S}_j (0) ) \nonumber \\
 && {\rm exp} \{ 
 \frac{i}{\hbar} (  I[{\bf S}_+ ]- I [{\bf S}_- ] ) 
\},  
\end{eqnarray}
where, $I_{qb} [I] = \int_0^t \hbar \Delta I_z$.
For following discussion, we choose the action of system,
$ I[ {\bf  S} ] = I_0 [ {\bf S} ] + I_{int} [ I_{z}, S_zi^0 ]$, thus the unperturbed is given by,
\begin{eqnarray}
   I_0 [ {\bf S} ] &=& \frac{ ( \gamma_N \hbar)^2}{4} 
   \int_0^t  \int_0^{t_1} dt_1 dt_2  
    A_{zz}^2 S_i^z (t_1)
    {\bf \Delta}_{00p}^{zz -1} ( t_1 , i , t_2 , i) 
    S_i^z ( t_2 ) \nonumber \\
   && \frac{1}{4} A_{\perp}^2 ( S_i^+ (t_1)
    {\bf \Delta}_{00p}^{+- -1} ( t_1 , i , t_2 , i) 
    S_i^- ( t_2 )
    + S_i^- (t_1)
    {\bf \Delta}_{00p}^{-+ -1} ( t_1 , i , t_2 , i) 
    S_i^+ ( t_2 )  )\nonumber \\
   &+& i \frac{\theta}{4\pi} \int dx \int_0^{t} dt \vec{n} 
        \cdot (\frac{\partial \vec{\bf{S}}}{\partial x} 
       \times \frac{\partial \vec{\bf{S}}}{\partial t} ) 
\end{eqnarray}
Where  $\theta=2 \pi n $ and ${\bf \Delta}_{00p} (t_1,i,t_2, j) $ is the free propagator of environmental system at zero temperature which is defined on closed time path and has four components,
\begin{eqnarray}
  \label{eqn:Green}
  && {\bf \Delta}_{00p}  (i, t_1, j, t_2) \nonumber \\ 
  &=& \left(
   \begin{array}{cc}
  {\bf \Delta}_{00}^{++} (i, t_1,j, t_2) & {\bf \Delta}_{00}^{+-}  
  (i, t_1, j, t_2) \\
  {\bf \Delta}_{00}^{-+}  (i, t_1,j, t_2) &{\bf  \Delta}_{00}^{--} 
  (i, t_1, j, t_2) 
  \end{array}
  \right). \nonumber \\
\end{eqnarray}
For the Berry phase term, because n is odd, the environment is a one-dimensional spin-1/2 system.
  \cite{Tsvelik}
In the present model, the spin chain is a gapless solvable model (a special point in spin chain systems).
Therefore, we can integrate over the spin chain degree of freedom and the Berry phase term does not make system to be distempered.
Introducing the incoming interaction picture for the environment system we can easily verify that Eq. (\ref{eqn:Green}) turns out,
\cite{Chou}
\begin{eqnarray}
  F[ I_{z+}, I_{z-}] &=& 
   {\rm exp} [ - i \frac{(\gamma_N \hbar)^2}{4} 
    \int_{0}^{t} \int_{0}^{t_1} dt_1 dt_2 \nonumber \\  
   &&  A_{zz}^2 
   ( I_{z+} (t_1) \Delta_{00}^{++} (i,t_1,i,t_2) I_{z+} (t_2) \nonumber \\
    &+& I_{z-} (t_1) \Delta_{00}^{--} (i,t_1,i,t_2) I_{z-} (t_2) \nonumber \\
    &-& I_{z+} (t_1) \Delta_{00}^{+-} (i,t_1,i,t_2) I_{z-} (t_2) \nonumber \\
    &-& I_{z-} (t_1) \Delta_{00}^{-+} (i,t_1,i,t_2) I_{z+} (t_2)) \nonumber \\ 
   &&  \frac{1}{4} A_{\perp}^2 
   ( I_{++} (t_1) \Delta_{00}^{++} (i,t_1,i,t_2) I_{++} (t_2) \nonumber \\
    &+& I_{+-} (t_1) \Delta_{00}^{--} (i,t_1,i,t_2) I_{+-} (t_2) \nonumber \\
    &-& I_{++} (t_1) \Delta_{00}^{+-} (i,t_1,i,t_2) I_{+-} (t_2) \nonumber \\
    &-& I_{+-} (t_1) \Delta_{00}^{-+} (i,t_1,i,t_2) I_{++} (t_2) )\nonumber \\
  && \frac{1}{4} A_{\perp}^2
   ( I_{-+} (t_1) \Delta_{00}^{++} (i,t_1,i,t_2) I_{-+} (t_2) \nonumber \\
   &+& I_{--} (t_1) \Delta_{00}^{--} (i,t_1,i,t_2) I_{--} (t_2) \nonumber \\
   &-&  I_{-+} (t_1) \Delta_{00}^{+-} (i,t_1,i,t_2) I_{--} (t_2) \nonumber \\
   &-& I_{--} (t_1) \Delta_{00}^{-+} (i,t_1,i,t_2) I_{-+} (t_2)) ]. \nonumber \\ \end{eqnarray}
For the convenience we change of the coordinate,
\begin{equation}
  \eta_z = (I_{z+} + I_{z-})/2, 
  \xi_z = (I_{z+} - I_{z-})/2.
\end{equation}
The $\eta$ and $\xi$ are called sojourn and blip.
In terms of this variable, density matrix is described by as follows
\begin{eqnarray}
 \rho (\eta_z =1) &=& |\uparrow><\uparrow|,
 \rho ( \eta_z =-1) = |\downarrow><\downarrow|, \nonumber \\
 \rho (\xi_z  = 1 ) &=&  |\uparrow><\downarrow|,
 \rho ( \xi_z = - 1)  = |\downarrow><\uparrow |. \nonumber 
\end{eqnarray}
Another variable are defined by
\begin{equation}
  \eta_+ = (I_{++} + I_{+-})/2, 
  \xi_+ = (I_{++} - I_{+-})/2,
\end{equation}
\begin{equation}
  \eta_- = (I_{-+} + I_{--})/2, 
  \xi_- = (I_{-+} - I_{--})/2.
\end{equation}
In terms of above variables the elements of density matrix are expressed as
\begin{eqnarray}
 \rho (\eta_+ =1) &=& |\uparrow><\downarrow|,
 \rho ( \eta_+ =-1) = |\downarrow><\uparrow|, \nonumber \\
 \rho (\xi_+  = 1 ) &=&  |\uparrow><\uparrow|,
 \rho ( \xi_+ = - 1)  = |\downarrow><\downarrow |. \nonumber \\
 \rho (\eta_- =1 ) &=& |\downarrow><\uparrow|,
 \rho ( \eta_- =-1) = |\uparrow><\downarrow|, \nonumber \\
 \rho (\xi_-  = 1 ) &=&  |\downarrow><\downarrow|,
 \rho ( \xi_- = - 1)  = |\uparrow><\uparrow |. \nonumber 
\end{eqnarray}
Therefore, the below equations are hold for these new variables,
\begin{equation}
\eta_z = \xi_+ = -\xi_- , \xi_z = \eta_+ =-\eta_-.
\end{equation}

In this case, the influence functional is expressed as
\begin{eqnarray}
 F[ \eta, \xi ] &=&
  {\rm exp} [ - i \frac{( \gamma_N \hbar)^2}{4}
 \int_{0}^{t} \int_{0}^{t_1} dt_1 dt_2  \nonumber \\
 && A_{zz}^2 
 \{ \xi (t_1)  G^R (t_1,i,t_2,i) \eta (t_2) \nonumber \\
 &+& \eta (t_1) G^A (t_1,i,t_2,i) \xi (t_2) \nonumber \\
 &-& \xi (t_1) G^K (t_1,i,t_2,i) 
                      \xi (t_2) \}   \nonumber \\
 &+& \frac{1}{2} ( 
A_{\perp}^2 \{ \eta (t_1)  G^R (t_1,i,t_2,i) \xi (t_2) \nonumber \\
 &+& \xi (t_1) G^A (t_1,i,t_2,i) \eta (t_2) \nonumber \\
 &-& \eta (t_1) G^K (t_1,i,t_2,i) 
                      \eta (t_2) \} )   \nonumber \\
 &+& \frac{1}{2} ( 
A_{\perp}^2 \{ \eta (t_1)  G^R (t_1,i,t_2,i) \xi (t_2) \nonumber \\
 &+& \xi (t_1) G^A (t_1,i,t_2,i) \eta (t_2) \nonumber \\
 &-& \eta (t_1) G^K (t_1,i,t_2,i) 
                      \eta (t_2) \} ) ]   \nonumber \\
&=&
  {\rm exp} [ - i \frac{( \gamma_N \hbar)^2}{4}
 \int_{0}^{t} \int_{0}^{t_1} dt_1 dt_2  \nonumber \\  
 && \{ \xi (t_1) ( A_{zz}^2 G^R (t_1,i,t_2,i) 
 +  A_{\perp}^2 G^A (t_1,i,t_2,i) )
  \eta (t_2) \nonumber \\
 &+& \eta (t_1) ( A_{zz}^2 G^A (t_1,i,t_2,i) 
 +  A_{\perp}^2 G^R (t_1,i,t_2,i) ) \xi (t_2)
                      \nonumber \\
                       &-& A_{zz}^2 \xi (t_1) G^K (t_1,i,t_2,i) 
                      \xi (t_2) \nonumber \\
 &-& 2 A_{zz}^2 \eta (t_1)  G^K (t_1,i,t_2,i) \eta (t_2) \}],
  \nonumber \\
\end{eqnarray}
where $G^R (t_1,i,t_2,j)$, $G^A (t_1,i,t_2,j)$ and $G^K (t_1,i,t_2,j)$ are retarded Green's function, advanced Green's function and Keldysh Green's function.
For above integral equation, after we slice the time and take a difference, we get the differential equation for density matrix is given by,
\begin{eqnarray}
&& \frac{ d \rho_{\rm b} (t)}{d t} = 
- \frac{i}{\hbar} [H_{qb}, \rho_{\rm b} (t)]
 -\frac{i}{4} (\gamma_N \hbar)^2 
\int_0^t dt_1\nonumber \\
&& \left(
 \begin{array}{cc}
  -A_{zz}^2 G_K (i, t,i, t_1), & A_{zz}^2 G^A (i, t,i, t_1) 
  + A_{\perp}^2 G^R (i, t,i, t_1)  \\
  A_{zz}^2 G^R (i, t,i, t_1) + A_{\perp}^2 G^A (i, t,i, t_1), 
  & - 2 A_{\perp}^2 G_K (i, t, i, t_1) \\ 
  \end{array}
  \right)  \rho_{\rm b} (t_1) \nonumber \\
\nonumber \\
\end{eqnarray}
where $[A,B]$ is $AB-BA$ and $\rho_{\rm b} (t)$ is
\begin{eqnarray}
\rho_{\rm b} (t) = \left( \begin{array}{cc}
  \eta_z (t) =1,  & \eta_z (t)= -1  \\
  \xi_z (t) =1, & \xi_z (t) = -1 \\ 
  \end{array} \right)
\end{eqnarray}  
Next we choose the representation of density matrix for the spin diagonal case,
\begin{eqnarray}
&& \frac{ d \rho_{\bf s} (t)}{d t}
= - \frac{i}{\hbar} [ H_{qb}, \rho_{\rm s}(t)] 
- \frac{i}{4} ( \gamma_N \hbar)^2 \int_0^t dt_1
\nonumber \\
&& \left(
  \begin{array}{c} 
- \frac{A_{zz}^2 +  A_{\perp}^2 }{2} G^K (i,t,i,t_1) \\
 \frac{A_{zz}^2 +  A_{\perp}^2}{2} ( G^R (i,t,i,t_1) + G^A (i,t,i,t_1)) \\
 \frac{i(A_{zz}^2 -  A_{\perp}^2 )}{2} 
 ( G^A (i,t,i,t_1) - G^R (i,t,i,t_1)) \\
 -\frac{ A_{zz}^2 -  A_{\perp}^2 }{2} G^K (i,t,i,t_1) \\
 \end{array}
 \right)^{t}  \rho_{\bf s} (t_1) \nonumber \\
\end{eqnarray}.
where the density matrix, $\rho_{\rm s} (t)$ is represented as
\begin{eqnarray}
\rho_{\rm s} (t) &=& \left( \begin{array}{c} 
  |\uparrow (t) >< \uparrow (t)| + |\downarrow(t)><\downarrow(t)|,\\
  |\uparrow (t)><\downarrow (t)|,  \\
  |\downarrow (t) >< \uparrow (t) |, \\ 
  |\uparrow (t) >< \uparrow (t)|-|\downarrow(t)><\downarrow(t)|  \\ 
  \end{array} \right). \nonumber \\
\end{eqnarray}  
The trace of density matrix decreases with time.
Another diagonal element show different behavior.
For spin flip process, another diagonal element increases with time, this represents self-excitation. 
The one of off-diagonal element become decoherence.
Another off-diagonal element shows oscillation where modulation of signal occurs.  

\section{one-dimensional qubits system}

Next, we examine 1-dimensonal qubit system by using influence functional.
\cite{Itakura}
In this case, the effect of indirect interaction appears.
The interaction Hamiltonian is as below,
\begin{equation}
  H_{int} = \gamma_N \hbar^2 \sum_i [  
  \frac{1}{2} 
A_{\perp} ( I_{i}^+ S_{i}^- + I_{i}^- S_{i}^+) +A_{zz}  I_{i}^z S_{i}^z ].
\end{equation}
We integrate about the qubit system except i-th site spin.
This lead density matrix of spin system,
\begin{eqnarray}
 \rho ( I_{zi+}^f, I_{zi-}^f )
   &=& \Pi_i \int^{I_{zi+}(t) =I_{zi+}^f, I_{zi-} (t) =
    I_{zi-}^f}_{ I_{zi+}(0)=I_{zi+}^f, I_{zi+} (0) = I_{zi+}^f} 
  [ d I_{zi+} ] [ d I_{zi-} ] \nonumber \\ 
 && {\exp} ( \frac{i}{\hbar} (I_{qb} [ I_{zi-} ] - I_{qb} [ I_{zi+} ] ))
  F[I_{zi+},I_{zi-}], 
\end{eqnarray}
the influence functional is
\begin{eqnarray}
 F[ I_{zi+}, I_{zj-} ] &=& \int [ d  {\bf S}_{+i}  ] [ d  {\bf S}_{-i} ]
 \delta (  {\bf S}_{+i} (t) - {\bf  S}_{-i}(t) ) \nonumber \\
 && \rho (  {\bf S}_i  (0),  {\bf S}_j (0) ) 
 {\rm exp} \{ 
 \frac{i}{\hbar} (  I[{\bf S}_+ ]- I [{\bf S}_- ] ) 
\}, \nonumber 
\end{eqnarray}
where, $I_{qb} [I] = \int_0^t \hbar \Delta I_iz$.
For following discussion, we choose the action of system,
$ I[ {\bf  S} ] = I_0 [ {\bf S} ] + I_{int} [ I_{zi}, S_zi^0 ]$, thus the unperturbed is given by,
\begin{eqnarray}
   I_0 [ {\bf S} ] &=& \frac{ (A_{zz} \gamma_N \hbar)^2}{4} 
   \int_0^t  \int_0^{t_1} dt_1 dt_2  \sum_{i.j} 
  {\bf S}_i^z (t_1)
    {\bf \Delta}_{00p}^{zz-1} ( t_1 , i , t_2 , j) 
    {\bf S}_j^z ( t_2 ) \nonumber \\
   &+& \frac{ (A_{\perp} \gamma_N \hbar)^2}{4} 
   ( {\bf S}_i^+ (t_1)
    {\bf \Delta}_{00p}^{+--1} ( t_1 , i , t_2 , j) 
    {\bf S}_j^- ( t_2 ) 
    +  {\bf S}_i^- (t_1)
    {\bf \Delta}_{00p}^{-+-1} ( t_1 , i , t_2 , j) 
    {\bf S}_j^+ ( t_2 ) \nonumber \\
   &+& i \frac{\theta}{4\pi} \int dx \int_0^{t} dt \vec{n} 
        \cdot (\frac{\partial \vec{\bf{S}}}{\partial x} 
       \times \frac{\partial \vec{\bf{S}}}{\partial t} ) 
\end{eqnarray}
where $\theta=2 \pi n $ and ${\bf \Delta}_{00p}^v (t_1,i,t_2, j) $ is the free propagator of environmental system at zero temperature which is defined on closed time path and has four components.
For the Berry phase term, because n is odd, the environment is a one-dimensional spin-1/2 system.
\cite{Tsvelik}
In the present model, the spin chain is a gapless solvable model (a special point in spin chain systems).
Therefore, we can completely integrate over the spin chain degree of freedom and the Berry phase term does not make system to be distempered.
Introducing the incoming interaction picture for the environment system we can easily verify, and the equation turns out,  
\begin{eqnarray}
  F[ I_{zi+}, I_{zi-}] &=& 
   {\rm exp} [ - i \frac{(\gamma_N \hbar)^2}{4} 
    \int_{0}^{t} \int_{0}^{t_1} dt_1 dt_2 \nonumber \\  
   &&  A_{zz}^2 
   ( I_{zi+} (t_1) \Delta_{00}^{++} (i,t_1,j,t_2) I_{zj+} (t_2) \nonumber \\
   &+& I_{zi-} (t_1) \Delta_{00}^{--} (i,t_1,j,t_2) I_{zj-} (t_2) \nonumber \\
    &-& I_{zi+} (t_1) \Delta_{00}^{+-} (i,t_1,j,t_2) I_{zj-} (t_2) \nonumber \\
  &-& I_{zi-} (t_1) \Delta_{00}^{-+} (i,t_1,j,t_2) I_{zj+} (t_2)) \nonumber \\ 
   &&  \frac{1}{4} A_{\perp}^2 
   ( I_{+i+} (t_1) \Delta_{00}^{++} (i,t_1,j,t_2) I_{+j+} (t_2) \nonumber \\
    &+& I_{+i-} (t_1) \Delta_{00}^{--} (i,t_1,j,t_2) I_{+j-} (t_2) \nonumber \\
    &-& I_{+i+} (t_1) \Delta_{00}^{+-} (i,t_1,j,t_2) I_{+j-} (t_2) \nonumber \\
   &-& 
   I_{+i-} (t_1) \Delta_{00}^{-+} (i,t_1,j,t_2) I_{+j+} (t_2) ) \nonumber \\
  && \frac{1}{4} A_{\perp}^2
   ( I_{-i+} (t_1) \Delta_{00}^{++} (i,t_1,j,t_2) I_{-j+} (t_2) \nonumber \\
    &+& I_{-i-} (t_1) \Delta_{00}^{--} (i,t_1,j,t_2) I_{-j-} (t_2) \nonumber \\
   &-&  I_{-i+} (t_1) \Delta_{00}^{+-} (i,t_1,j,t_2) I_{-j-} (t_2) \nonumber \\
   &-& I_{-i-} (t_1) \Delta_{00}^{-+} (i,t_1,j,t_2) I_{-j+} (t_2)) ].  
\end{eqnarray}
For the convenience we change of the coordinate,
\begin{equation}
  \eta_{zi} = (I_{zi+} + I_{zi-})/2, 
  \xi_{zi} = (I_{zi+} - I_{zi-})/2.
\end{equation}
The $\eta$ and $\xi$ are called sojourn and blip.
In terms of this variable, density matrix is described by as follows
\begin{eqnarray}
 \rho (\eta_{zi} =1) &=& |\uparrow_i><\uparrow_i|,
 \rho ( \eta_{zi} =-1) = |\downarrow_i><\downarrow_i|, \nonumber \\
 \rho (\xi_{zi}  = 1 ) &=&  |\uparrow_i><\downarrow_i|,
 \rho ( \xi_{zi} = - 1)  = |\downarrow_i><\uparrow_i |. \nonumber 
\end{eqnarray}
Another variable are defined by
\begin{equation}
  \eta_{+i} = (I_{+i+} + I_{+i-})/2, 
  \xi_{+i} = (I_{+i+} - I_{+i-})/2,
\end{equation}
\begin{equation}
  \eta_{-i} = (I_{-i+} + I_{-i-})/2, 
  \xi_{-i} = (I_{-i+} - I_{-i-})/2.
\end{equation}
In terms of above variables the elements of density matrix are expressed as
\begin{eqnarray}
 \rho (\eta_{+i} =1) &=& |\uparrow_i><\downarrow_i|,
 \rho ( \eta_{+i} =-1) = |\downarrow_i><\uparrow_i|, \nonumber \\
 \rho (\xi_{+i}  = 1 ) &=&  |\uparrow_i><\uparrow_i|,
 \rho ( \xi_{+i} = - 1)  = |\downarrow_i><\downarrow_i |. \nonumber \\
 \rho (\eta_{-i} =1 ) &=& |\downarrow_i><\uparrow_i|,
 \rho ( \eta_{-i} =-1) = |\uparrow_i><\downarrow_i|, \nonumber \\
 \rho (\xi_{-i}  = 1 ) &=&  |\downarrow_i><\downarrow_i|,
 \rho ( \xi_{-i} = - 1)  = |\uparrow_i><\uparrow_i |. \nonumber 
\end{eqnarray}
Therefore, the below equations are hold for these new variables,
\begin{equation}
\eta_{zi} = \xi_{+i} = -\xi_{-i} , \xi_{zi} = \eta_{+i} =-\eta_{-i}.
\end{equation}

In this case, the influence function is expressed as
\begin{eqnarray}
&=&
  {\rm exp} [ - i \frac{( \gamma_N \hbar)^2}{4}
 \int_{0}^{t} \int_{0}^{t_1} dt_1 dt_2 \nonumber \\  
 && \{ \xi_i (t_1) ( A_{zz}^2 G^R (t_1,i,t_2,j) 
 + A_{\perp}^2 G^A (t_1,i,t_2,j) )
  \eta_j (t_2) \nonumber \\
 &+& \eta_i (t_1) ( A_{zz}^2 G^A (t_1,i,t_2,j) 
 + A_{\perp}^2 G^R (t_1,i,t_2,j) ) \xi_j (t_2)
                      \nonumber \\
                       &-& A_{zz}^2 \xi_i (t_1) G^K (t_1,i,t_2,j) 
                      \xi_j (t_2) 
      - 2 A_{zz}^2 \eta_i (t_1)  G^K (t_1,i,t_2,j) \eta_j (t_2) \}],
  \nonumber \\
\end{eqnarray}
where $G^R (t_1,i,t_2,j)$, $G^A (t_1,i,t_2,j)$ and $G^K (t_1,i,t_2,j)$ are retarded Green's function, advanced Green's function and Keldysh Green's function.
For above integral equation, after slice the time and take difference, we get the differential equation for density matrix is given by,
\begin{eqnarray}
&&\frac{ d \rho_{\rm b}  (i,j,t)}{d t} = 
- \frac{i}{\hbar} [H_{qb}, \rho_{\rm b}(t)] 
-\frac{i}{4} (\gamma_N \hbar)^2 
\int_0^t dt_1 \sum_k  \nonumber \\
&& \left(
 \begin{array}{cc}
  -A_{zz}^2 G_K (i, t,k, t_1) & A_{zz}^2 G^A (i, t,k, t_1) 
  +  A_{\perp}^2 G^R (i, t,k, t_1)  \\
  A_{zz}^2 G^R (i, t,k, t_1) + A_{\perp}^2 G^A (i, t,k, t_1) 
  & - A_{\perp}^2 G_K (i, t, k, t_1) \\ 
  \end{array}
  \right) \rho_{\rm b} (t_1,k,j). \nonumber \\
\end{eqnarray}
Next we choose the representation of density matrix for the spin diagonal case,
\begin{eqnarray}
&& \frac{ d \rho_{\rm s} (i,j,t)}{d t} = 
 - \frac{i}{\hbar}
  [H_{qb}, \rho_{\rm s} (t)] - \frac{i}{4} ( \gamma_N \hbar)^2 \int_0^t dt_1 \sum_k
 \nonumber \\
&& \left(
  \begin{array}{c} 
- \frac{A_{zz}^2 + A_{\perp}^2 }{2} G^K (i,t,k,t_1) \\
 \frac{A_{zz}^2 + A_{\perp}^2}{2} ( G^R (i,t,k,t_1) + G^A (i,t,k,t_1)) \\
 \frac{i(A_{zz}^2 - A_{\perp}^2 )}{2}
  ( G^A (i,t,k,t_1) - G^R (i,t,k,t_1)) \\
 -\frac{ A_{zz}^2 - A_{\perp}^2 }{2} G^K (i,t,k,t_1) \\
 \end{array}
 \right) \rho_{\rm s} (k,j,t_1). \nonumber \\
\end{eqnarray}
The trace of density matrix decreases with time.
Another diagonal element shows different behavior.
For spin flip process, another diagonal element increases with time, this represents self-excitation. 
The off-diagonal element shows oscillation where modulation of the signal occurs.
The self-excitation is appears in poor man scale for Kondo effect.
The strong coupling limit is self-excitation.
Above behavior is simple RG flow.

We examine the pure dephasing event.
Because the propagator of qubit has no time dependence, the blip state and sojourn state does not change.
Therefore when we choose the initial condition of qubit density matrix to be coherent state, such as, 
\begin{eqnarray}
   \rho ( t=0) = \prod_i
                \pm ( |\uparrow_i><\downarrow_i| 
                \pm |\downarrow_i><\uparrow_i|) ,
                \nonumber \\
\end{eqnarray}
the time evolution occurs only off-diagonal channel.
In addition to that, even if we start from off-diagonal state, the interaction Hamiltonian does not contain the spin flip process.
Then the blip state does not change.
So we can take the $\xi_i (t)=\xi_i (=\pm 1)$ and $\eta_i (t)=0$ for all t.
This situation leads to exact expression of dephasing rate as follows,
\begin{eqnarray} 
 && \rho (\xi_i,t) = \int^{I_{zi+}(t) =I_{zi+}^f, I_{zi-} (t) =
    I_{zi-}^f}_{ I_{zi+}(0)=I_{zi+}^f, I_{zi+} (0) = I_{zi+}^f} 
  d  \xi_i  \nonumber \\
 && {\rm exp} [ i \Delta t -i 
  \frac{( A_{zz} \gamma_N \hbar )^2 }{4} 
  \sum_{i \ne j} \int_0^t dt \int_0^{t_1} dt_2 
  \xi_i G^K (t_1,i,t_1,j) \xi_j ] \nonumber \\ 
 &=& {\rm exp} ( i \Delta t ) 
 \times  
 {\rm det}_i [ -  i \frac{(A_{zz} \gamma_N \hbar)^2 }{16} 
 \int^{t}_0 dt_1 \int^{t_1}_0 dt_2
 G^K (t_1,i,t_2,j)  ] \nonumber \\ 
 &=& {\rm exp} [ i \Delta t  
 - {\rm Tr} 
 \log [ 1 - i \frac{(A_{zz} \gamma_N \hbar)^2 }{16}
  \int^{t}_0 dt_1 \int^{t_1}_0 dt_2 
 G^K (t_1,i,t_2,i)] ] , \nonumber \\ 
\end{eqnarray}
where we neglect the effect of direct interaction.
Using the analytic expression, the time evolution of off diagonal density matrix element is given by
\begin{eqnarray}
  &-& \ln \Re \rho(\xi_i = \pm 1,t ) \nonumber \\
 &=& {\rm Tr}   
  \log [ 1 + \frac{( A_{zz} 
  \gamma_N \hbar)^2 }{16}
   \int_0^t dt \int_0^{t_1} dt_2 
   \{ S^z_j (t_1), S^z_j (t_2) \}  ] \nonumber \\
 &=& 
  L \log [ 1 + \frac{( A_{zz} 
  \gamma_N \hbar)^2 }{16} 
  \int_0^t dt \int_0^{t_1} dt_2 
       \{ S^z_0 (t_1), S^z_0 (t_2) \}  ] \nonumber \\
  &=& L  \log [ 1 + \frac{( A_{zz} \gamma_N \hbar)^2}{16} 
  \int_{-\infty}^{\infty} dw 
     \{ S^z_0 (w), S^z_0 (-w) \}  
   \left(
   \frac{\sin( \omega t / 2)}{\omega/2} \right)^2  ] 
   \nonumber \\
\end{eqnarray}
where $ \{ S^z_0,S^z_0 \} $ is symmetries correlation function, defined by $\{ A,B \} =AB + BA$ because Keldysh Green's function is symmetries correlation function  for the base of present Hilbert space.
Then,
\begin{equation}
\rho_{\rm off} (t)
 = (1+\frac{(A_{zz} \gamma_N \hbar)^2}{16} \int_{-\infty}^{\infty} dw 
     \{ S^z_0 (w), S^z_0 (-w) \}  
   \left(
   \frac{\sin( \omega t / 2)}{\omega/2} \right)^2 )^{-L}
\end{equation}
Above result is exact.
Thus, for infinite number of qubit system the dephasing rate becomes infinity.
Next we evaluate the quality factor.
The quantum error correction code rate is given by $\Delta L$,
where $\Delta$ is single qubit coherence time and $L$ is number of qubit.
The decoherence rate is given by $\ln (1+  \frac{T_{2}^{-1}}{4} t) \frac{ L}{t} $
where $T_{2}^{-1}$ is decoherence rate by single qubit.
Therefore the quality factor is given by 
$Q=\frac{\Delta t }{ \ln(1+ \frac{T_{2}^{-1} }{4} t ) }$.
At $t \rightarrow \infty$, $Q$ becomes infinity, thus spin quantum computer is scalable.

\section{Numerical results}
For Gaussian noise, we obtain the numerical results.
The initial conditions are $I(t),\sigma_x,\sigma_y,\sigma_z=1$.
The coupling constants are $A_{zz}=1.0$, $A_{\perp}=0.1$.
The off-diagonal components show oscillation.
The trace shows decoherence.
The diagonal component increases.
The purity shows oscillation.
The definition of purity is given by $Tr (\rho^2(t))$

For $SU(2)$ symmetric case, $(A_{zz} = A_{\perp})$,
the equation for density matrix is given by as follows,
\begin{eqnarray}
\frac{d I (t)}{d t} &=& - \frac{1}{4} (\gamma_N \hbar A)^2 \int^t_0 dt_1 
C(t-t_1) \sigma_{z} \nonumber \\
\frac{d \sigma_z (t)}{ d  t} &=& 
-i \frac{\Delta}{\hbar} \sigma_y (t) + \frac{1}{4} ( \gamma_N \hbar)^2 \int_0^t
dt_1 C(t-t_1) \sigma_{x} (t_1) \nonumber \\
\frac{d \sigma_y (t)}{d t} &=& -i \frac{\Delta}{\hbar} (t) \nonumber \\
\frac{d \sigma_z (t)}{dt} &=& 0.
\end{eqnarray}
By using above equations, the equation for $\sigma_y (t)$ is
\begin{eqnarray}
\frac{d^2 \sigma_y (t)}{d t^2} &=& - \frac{\Delta^2}{\hbar^2} \sigma_{y} (t)
+ i \frac{\hbar}{4 \Delta} (\gamma_N A)^2 \int_0^t dt_1 C(t-t_1) \frac{d \sigma_y (t)}{d t}
\end{eqnarray}
The numerical results for $SU(2)$ symmetric case is ginven as Fig.6-10.

\begin{figure}[tb]
\begin{center}
\unitlength 1mm
\begin{picture}(90,90)(0,5)
\put(0,0){\resizebox{65mm}{!}{\includegraphics{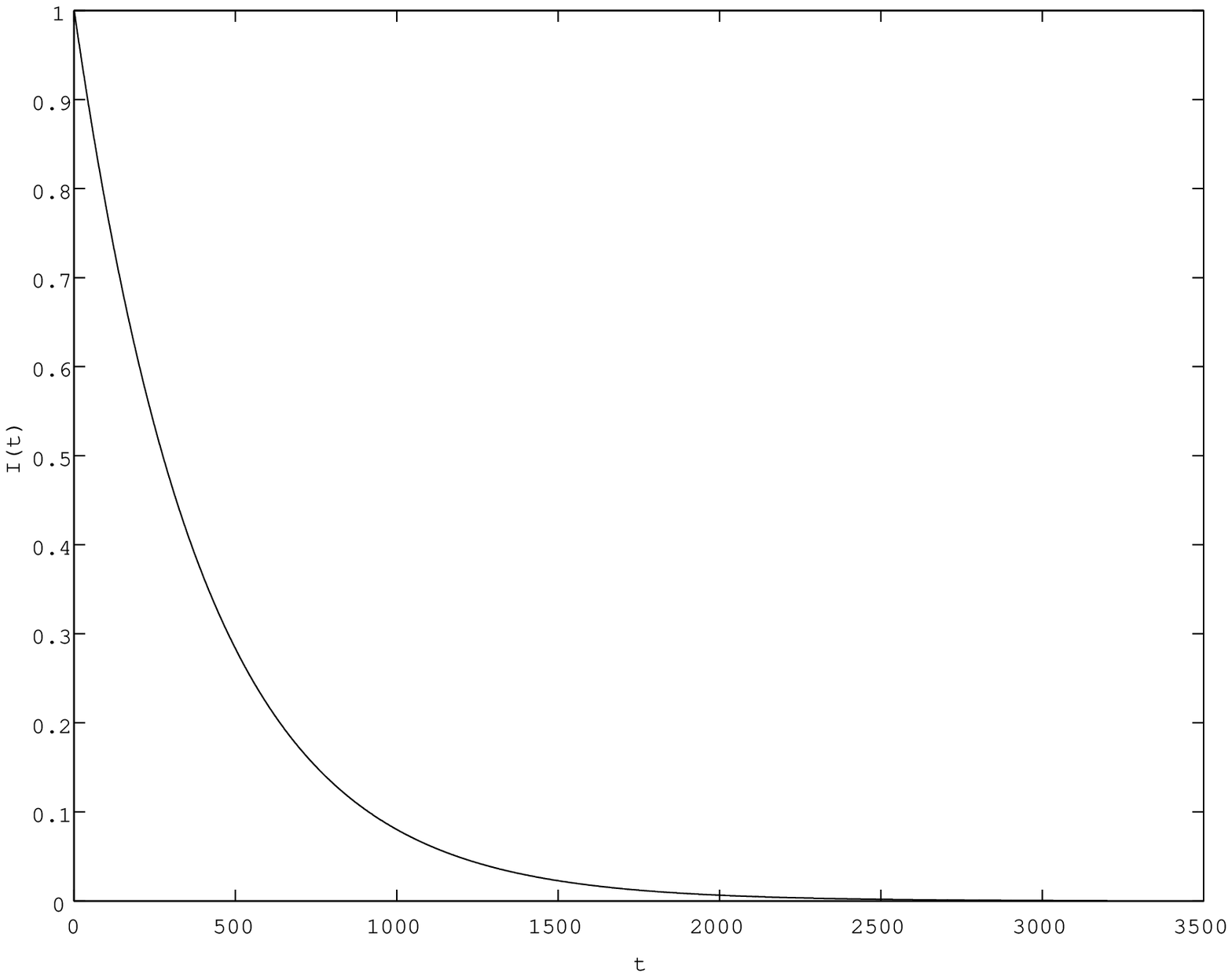}}}
\end{picture}
\end{center}
\caption{\label{fig:exact} Real part of $I(t)$}
\end{figure}
\begin{figure}[tb]
\begin{center}
\unitlength 1mm
\begin{picture}(90,90)(0,5)
\put(0,0){\resizebox{65mm}{!}{\includegraphics{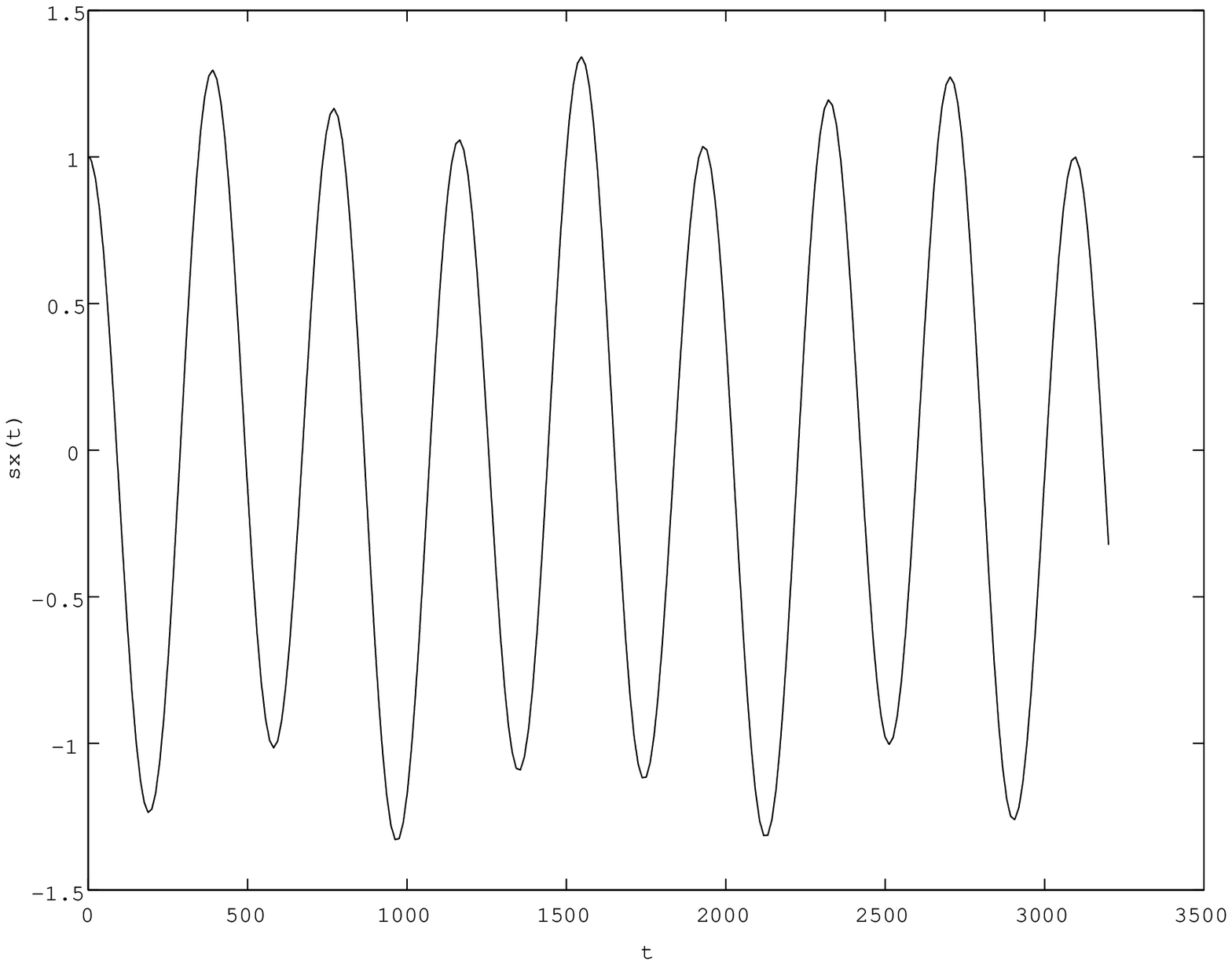}}}
\end{picture}
\end{center}
\caption{\label{fig:exact} Real part of $\sigma_x$}
\end{figure}
\begin{figure}[tb]
\begin{center}
\unitlength 1mm
\begin{picture}(90,90)(0,5)
\put(0,0){\resizebox{65mm}{!}{\includegraphics{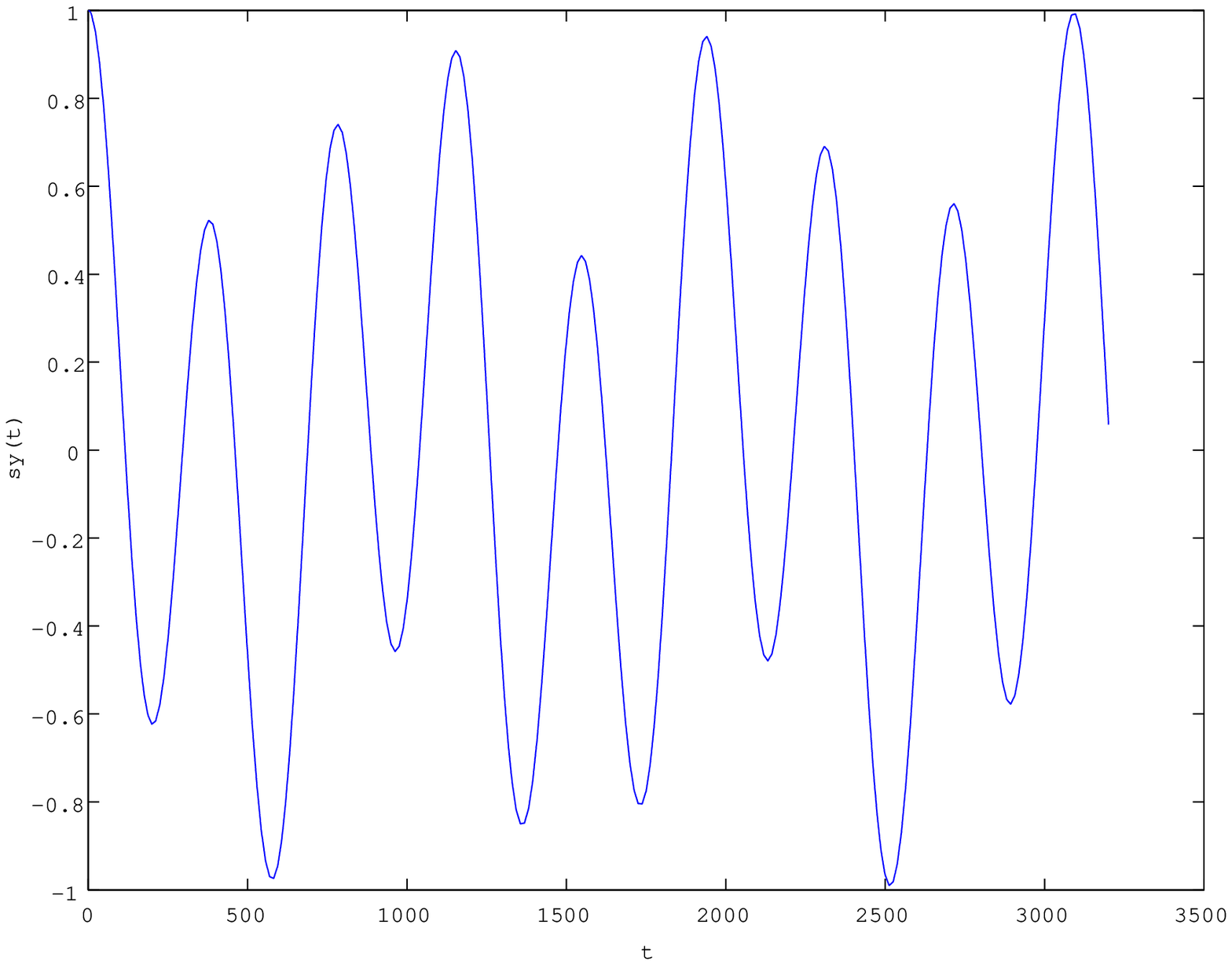}}}
\end{picture}
\end{center}
\caption{\label{fig:exact} Real part of $\sigma_y$}
\end{figure}
\begin{figure}[tb]
\begin{center}
\unitlength 1mm
\begin{picture}(90,90)(0,5)
\put(0,0){\resizebox{65mm}{!}{\includegraphics{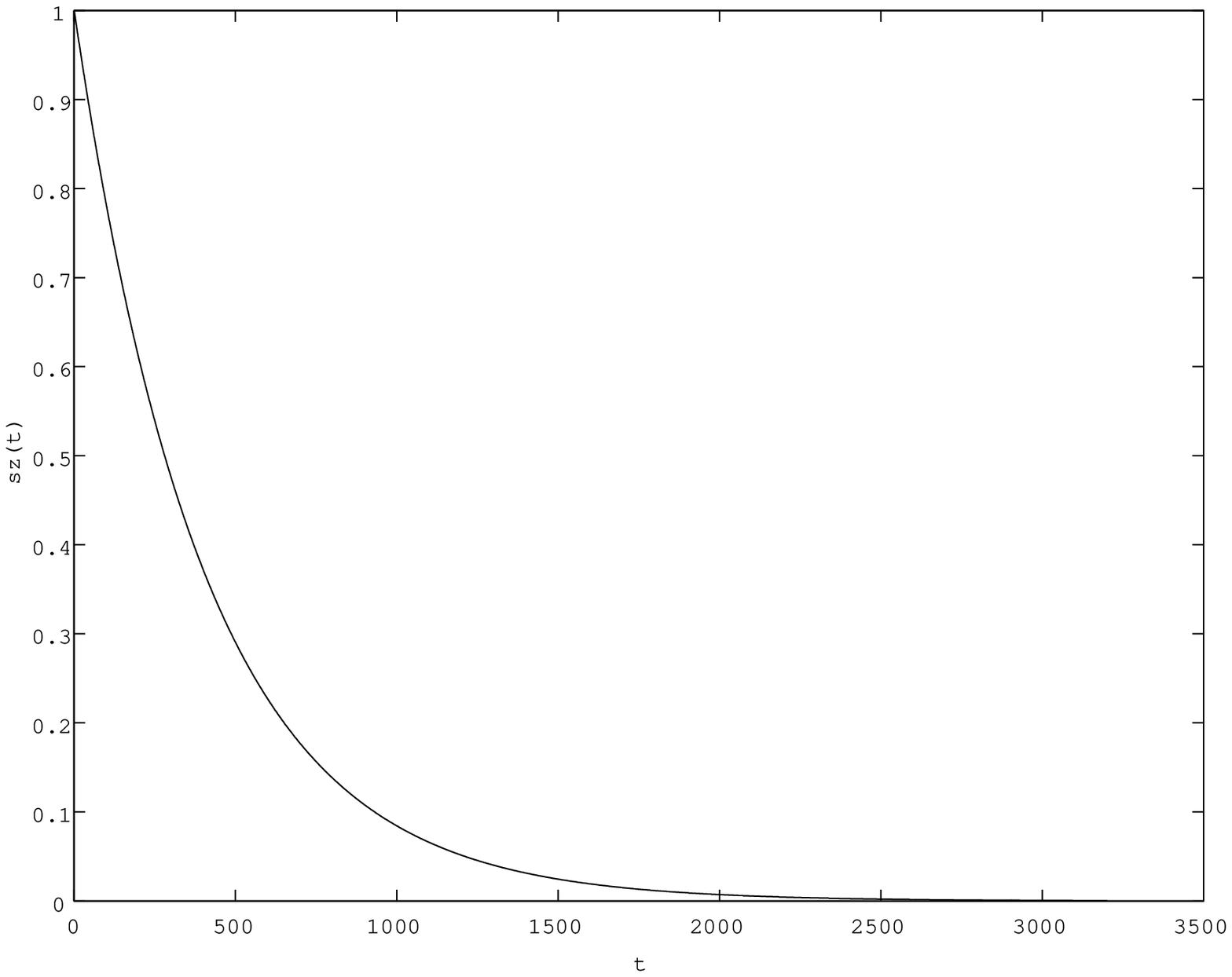}}}
\end{picture}
\end{center}
\caption{\label{fig:exact} Real part of $\sigma_z$}
\end{figure}
\begin{figure}[tb]
\begin{center}
\unitlength 1mm
\begin{picture}(90,90)(0,5)
\put(0,0){\resizebox{65mm}{!}{\includegraphics{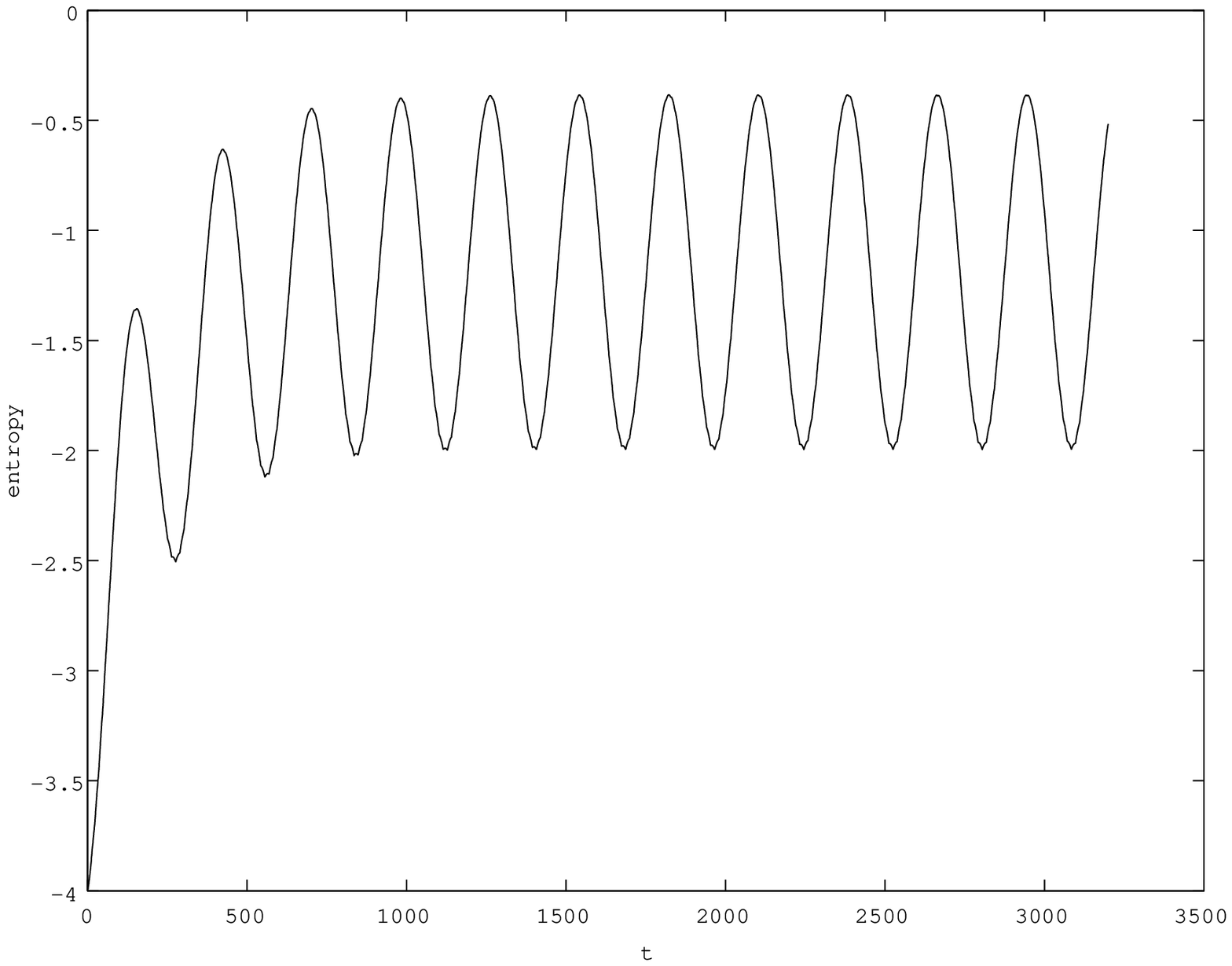}}}
\end{picture}
\end{center}
\caption{\label{fig:exact} purity}
\end{figure}
\begin{figure}[tb]
\begin{center}
\unitlength 1mm
\begin{picture}(90,90)(0,5)
\put(0,0){\resizebox{65mm}{!}{\includegraphics{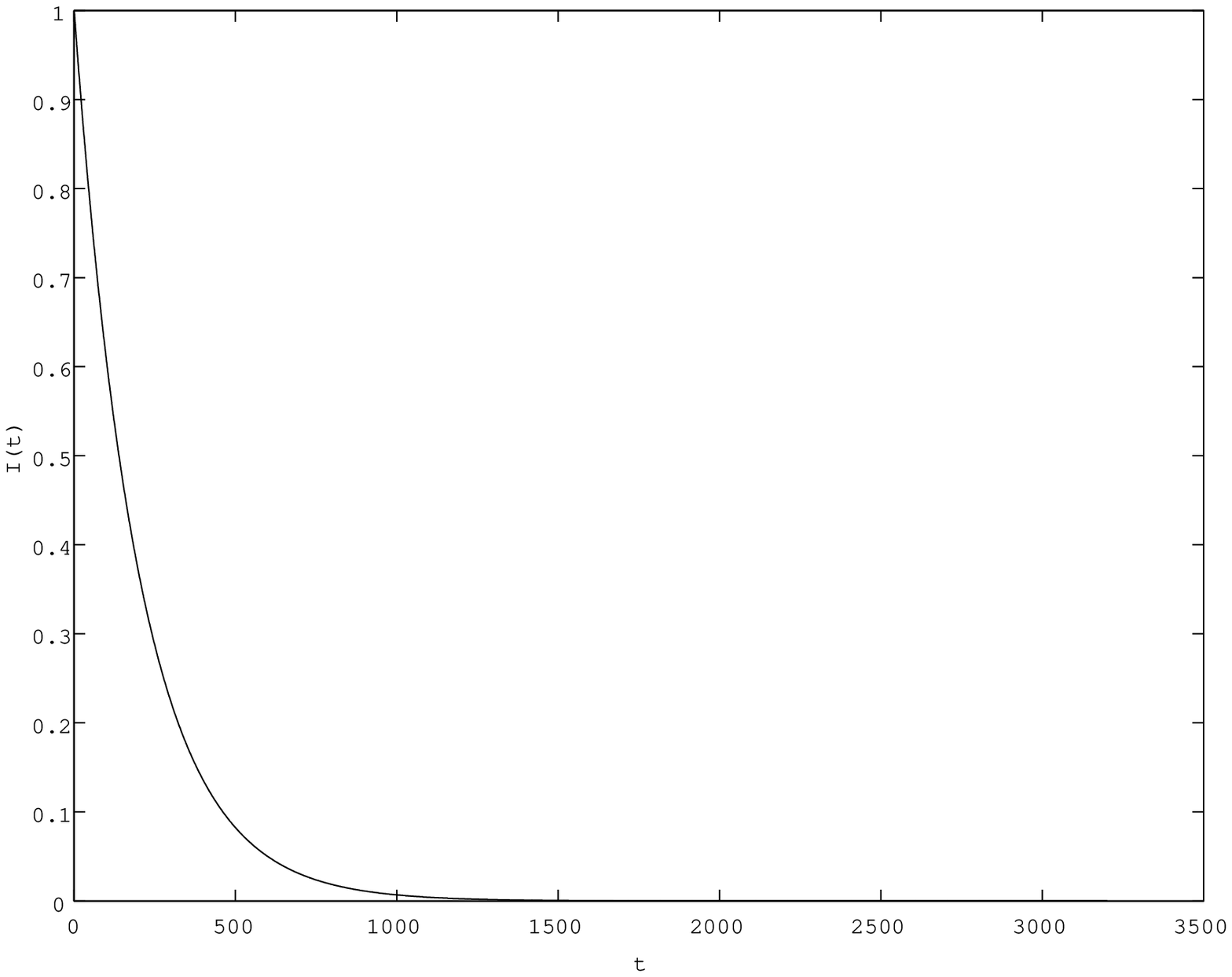}}}
\end{picture}
\end{center}
\caption{\label{fig:exact} Real part of $I(t)$}
\end{figure}
\begin{figure}[tb]
\begin{center}
\unitlength 1mm
\begin{picture}(90,90)(0,5)
\put(0,0){\resizebox{65mm}{!}{\includegraphics{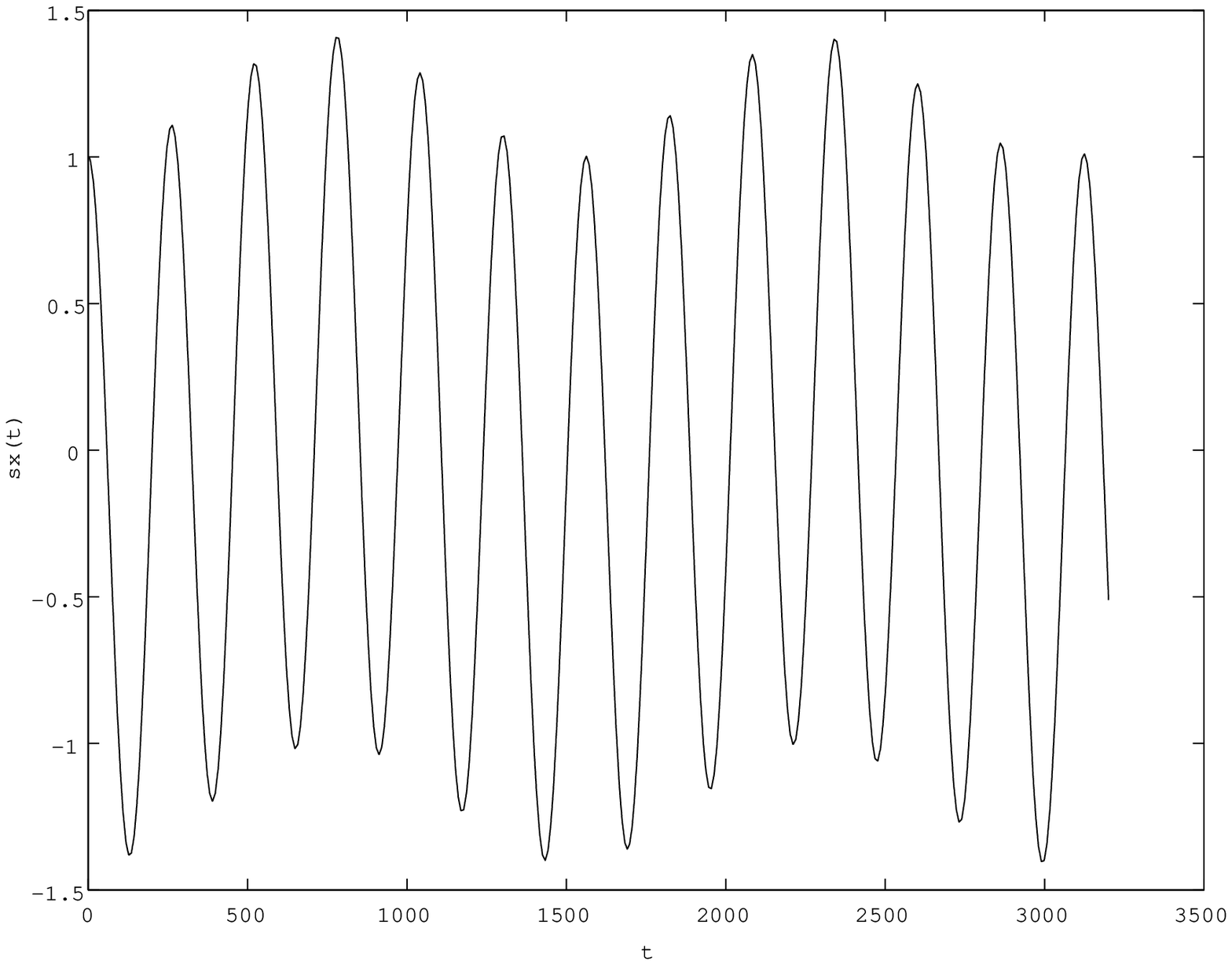}}}
\end{picture}
\end{center}
\caption{\label{fig:exact} Real part of $\sigma_x$}
\end{figure}
\begin{figure}[tb]
\begin{center}
\unitlength 1mm
\begin{picture}(90,90)(0,5)
\put(0,0){\resizebox{65mm}{!}{\includegraphics{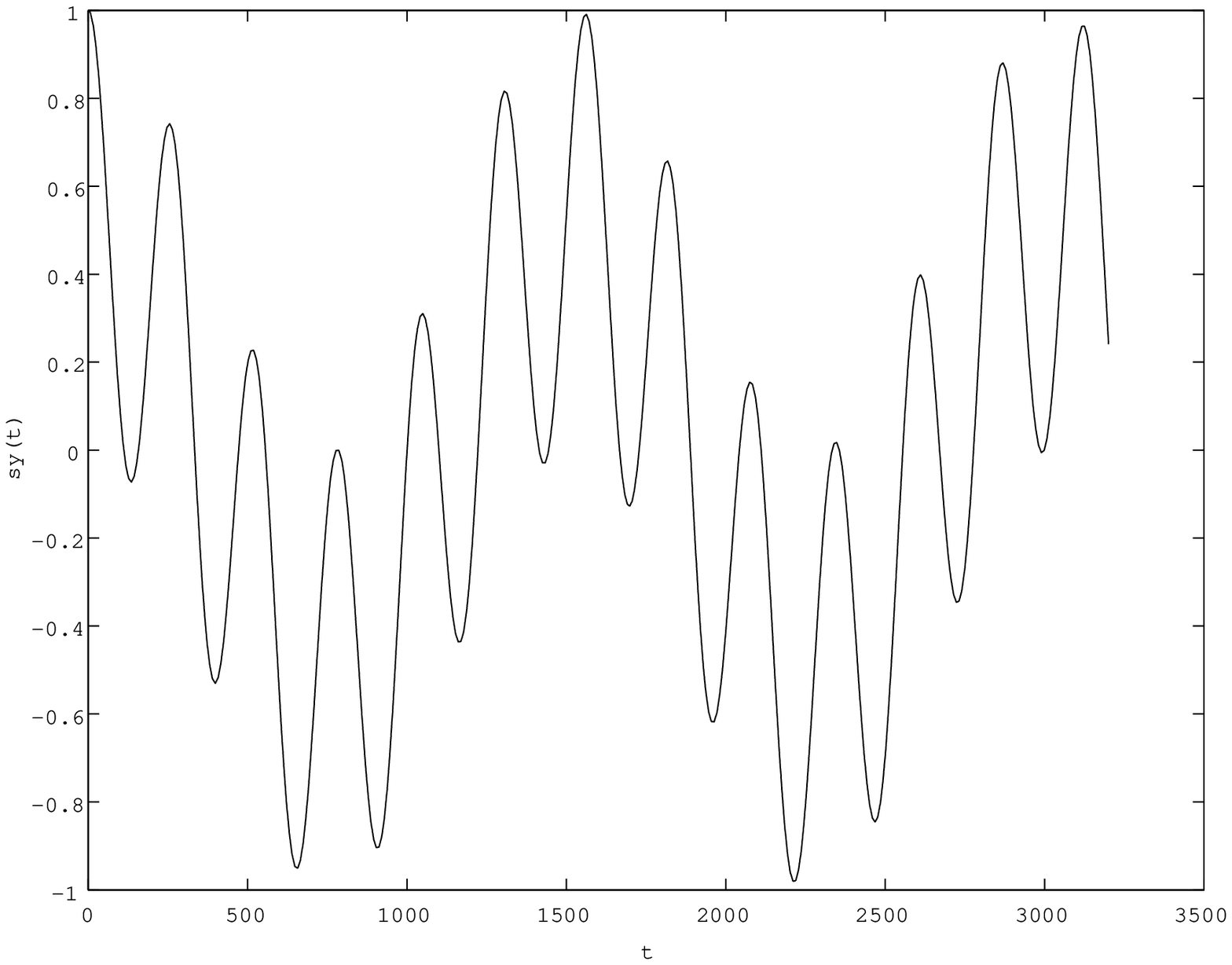}}}
\end{picture}
\end{center}
\caption{\label{fig:exact} Real part of $\sigma_y$}
\end{figure}
\begin{figure}[tb]
\begin{center}
\unitlength 1mm
\begin{picture}(90,90)(0,5)
\put(0,0){\resizebox{65mm}{!}{\includegraphics{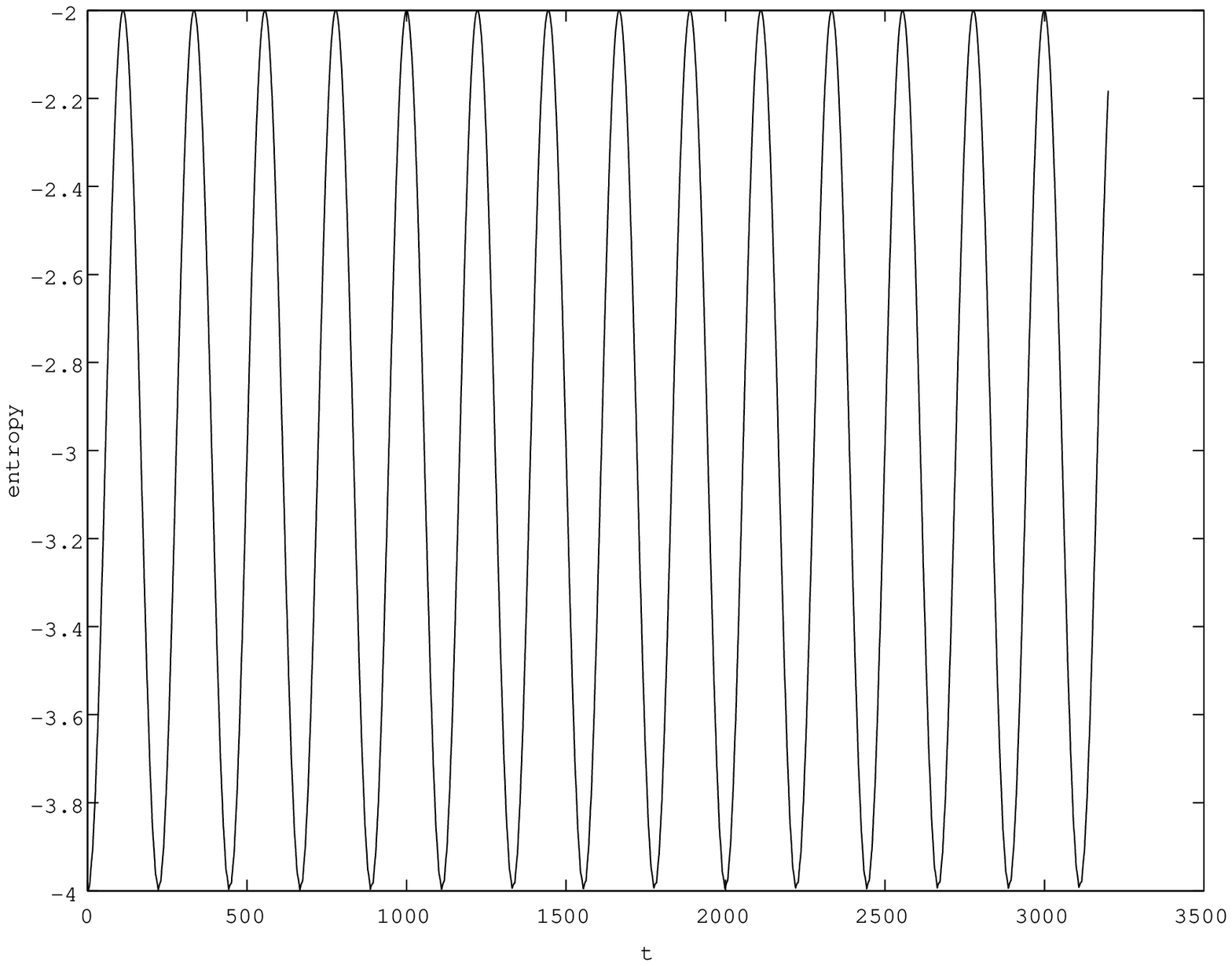}}}
\end{picture}
\end{center}
\caption{\label{fig:exact} purity}
\end{figure}

\section{Conclusion}

In summary, we had examined decoherence of a spin qubits system coupled with a spin chain.
The examined decoherence is the case of single qubit system and one-dimensional qubits system.
We obtained the differential-integral equation for general initial condition.
The trace of density matrix decreases with time.
Another diagonal element shows different behavior.
For spin flip process, another diagonal element increases with time, this represents self-excitation. 
The off-diagonal element shows oscillation where modulation of the signal occurs.
Decoherence without trace conservation occurs. 
At thermodynamic limit, quality factor becomes infinity,
 thus spin quantum computer is scalable.
We also examine the numerical calculation for Gaussian noise.
The purity is plotted, this quantity shows oscillating behavior.


\end{document}